\documentclass[prl,reprint,aps,twocolumn]{revtex4-1}

\pdfoutput=1

\usepackage[utf8]{inputenc}

\usepackage[T1]{fontenc}
\usepackage{hyperref}
\usepackage{color}
\usepackage{graphicx}
\usepackage{units}
\usepackage{amsmath}

\usepackage{epstopdf}
\usepackage{subcaption}
\usepackage{amssymb}
\usepackage{pifont}
\usepackage{textcomp}
\usepackage[usenames,dvipsnames]{xcolor}
\usepackage{ulem}
\usepackage{amsmath} 

\newcommand{\SI}[1]{{#1}}

\hypersetup{colorlinks = true, citecolor = blue, breaklinks = true}

\begin{document}

\title{Anion order and spontaneous polarization in LaTiO\textsubscript{2}N oxynitride thin films}
\date{\today}
\author{Nathalie Vonr\"uti}
\affiliation{Department of Chemistry and Biochemistry, University of Bern, Freiestrasse 3, CH-3012 Bern, Switzerland}
\author{Ulrich Aschauer}
\affiliation{Department of Chemistry and Biochemistry, University of Bern, Freiestrasse 3, CH-3012 Bern, Switzerland}

\begin{abstract}
The perovskite oxynitride LaTiO\textsubscript{2}N is a promising material for photocatalytic water splitting under visible light. 
One of the obstacles towards higher efficiencies of this and similar materials stems from charge-carrier recombination, which could be suppressed by the built-in electric field in polar materials. In this study, we investigate the spontaneous polarization in epitaxially 
strained LaTiO\textsubscript{2}N thin films via density functional theory calculations. The effect of epitaxial strain on the anion order, resulting out-of-plane polarization, 
energy barriers for polarization reversal and the corresponding coercive fields are studied. We find that for compressive strains larger than 4\% the thermodynamically stable anion order is polar along the 
out-of-plane direction and has a coercive field comparable to other switchable ferroelectrics. Our results show that 
strained LaTiO\textsubscript{2}N could indeed suppress carrier recombination and lead to enhanced photocatalytic activities.
\end{abstract}

\maketitle
Complex oxides, for example in the perovskite structure, represent a promising class of materials for photocatalytic water-splitting electrodes. 
Compared to binary oxides, such as the prototypical TiO\textsubscript{2} \cite{fujishima1972electrochemical}, their chemical and structural flexibility 
allows tuning the electronic band structure to decrease the band gap and hence extend light absorption from the ultraviolet to the visible part of the solar spectrum. 
A further reduction of the band gap results from partial substitution of oxygen ions 
with less electronegative nitrogen ions in so-called perovskite oxynitrides \cite{ebbinghaus2009perovskite, takata2015recent, ahmed2016review, wu2013first, castelli2012new}. 
Due to the higher-lying N 2p states forming the valence-band edge, these materials can have band gaps just above 2 eV \cite{Kasahara:2002kd}, resulting in a good visible-light photo-absorption efficiency, 
unfortunately at the cost of a slightly reduced chemical stability \cite{Fuertes2010}. 

One source for low photocatalytic efficiencies, not restricted to oxynitrides, is the recombination of photo-induced electron-hole pairs before reaching reactive surface sites \cite{yamasita2004recent}.
It was previously shown that the built-in electric field in polar materials significantly enhances charge-carrier separation  \cite{giocondi2001spatial}, as electrons and holes migrate in opposite directions. 
In ferroelectric materials, switching of the polarization direction by an applied electric field binds and releases molecules from the surface, turning reaction pathways on and off to enable a new level of control in heterogenous catalysis \cite{garrity2010chemistry, garrity2013ferroelectric}. 
Most importantly, this switching could offer a way to go beyond Sabatier's principle, which states that the tradeoff between strong reactant adsorption and facile product desorption limits the overall performance of a catalyst. Cyclic switching could enable strong product adsorption in one polarization state while products are released easily in the oppositely poled state \cite{Kakekhani:2015ek}.

Epitaxial strain is an established way to trigger or enhance spontaneous polarization in complex perovskite oxides \cite{haeni2004room, Choi:2004it}. This can be understood in the second-order Jahn-Teller picture \cite{Rondinelli:2009tq, Bersuker:2013fa}, where the existence of a polar instability is determined by a competition between a coulombic and a covalent energy contribution. The former is positive and favors the centrosymmetric structure, while the latter is negative and favors the polar structure. Elongation of a crystal reduces the coulomb repulsion leading to the appearance of the polar distortion as the covalent term starts to dominate \cite{aschauer2014competition}. With epitaxial strain, an elongation occurs either biaxially in-plane for tensile or uniaxially out-of-plane for compressive strain.

Compared to pure oxides, strain effects are more complex in oxynitrides as not only the cell shape and ion positions but also the anion order is affected. Most bulk oxynitrides do not show long-range anion order, while locally a \textit{cis} coordination, which maximizes the overlap 
of 2p and d orbitals, is often preferred \cite{attfield2013principles,yang2011anion, clarke2002oxynitride, fuertes2012chemistry}. Under compressive epitaxial strain nitrogen rather than oxygen tends to occupy sites along the out-of-plane direction in Ca\textsubscript{1-x}Sr\textsubscript{x}TaO\textsubscript{2}N \cite{oka2014possible,oka2017strain}. Other perovskite oxynitrides such as YTiO$_2$N also have this \textit{trans}-type order and adopt polar structures \cite{Caracas:2007el}.

\begin{figure}
\centering
\includegraphics[width=0.9\columnwidth]{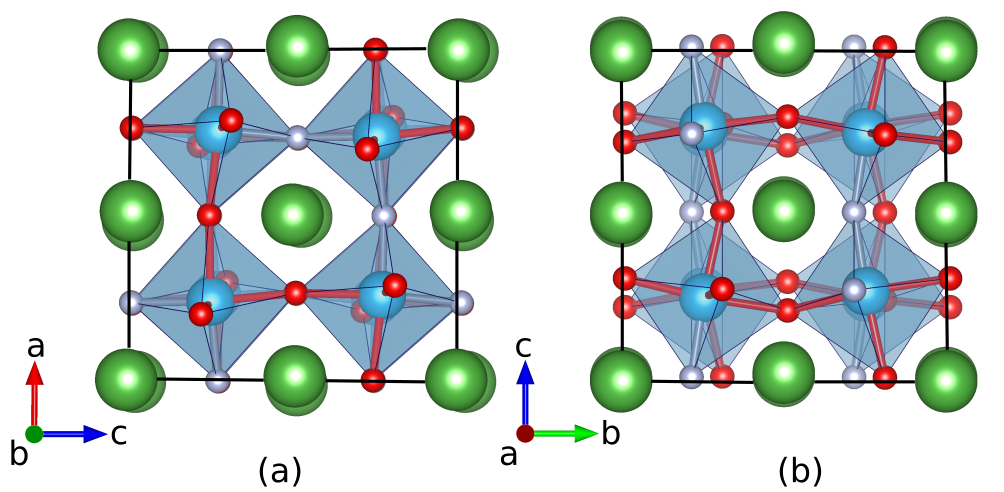}
\caption{(a) Side and (b) top view of the thermodynamically most stable unstrained LaTiO\textsubscript{2}N structure with \textit{cis}-anion order \cite{ninova2017surface}. Color code: La=green,Ti=blue,O=red,N=grey.}
\label{fig:cell}
\end{figure}

To asses the prospect of epitaxial strain to enhance the photocatalytic activity of oxynitrides, we investigate in this letter, using density functional theory (DFT) calculations, the strain-dependent thermodynamic stability and spontaneous polarization of different anion orders in the well-studied perovskite oxynitride LaTiO\textsubscript{2}N shown in Fig. \ref{fig:cell}. Our DFT calculations are performed at the PBE+U level of theory \cite{perdew1996generalized, anisimov1991band} using the Quantum ESPRESSO package \cite{giannozzi2009quantum}. Further computational details can be found in the \SI{supporting information}. We start from the \textit{cis}-ordered structure \cite{ninova2017surface} shown in Fig. \ref{fig:cell}, which has space group P1, a\textsuperscript{-}b\textsuperscript{+}c\textsuperscript{-} octahedral rotations in Glazer notation \cite{Glazer:1972eb} and, in line with other experimental and theoretical studies \cite{yashima2010imma, wang2015hybrid, kasahara2003latio2n}, is thermodynamically stable in the unstrained bulk phase. 

\begin{figure}
\centering
\includegraphics[width=0.9\columnwidth]{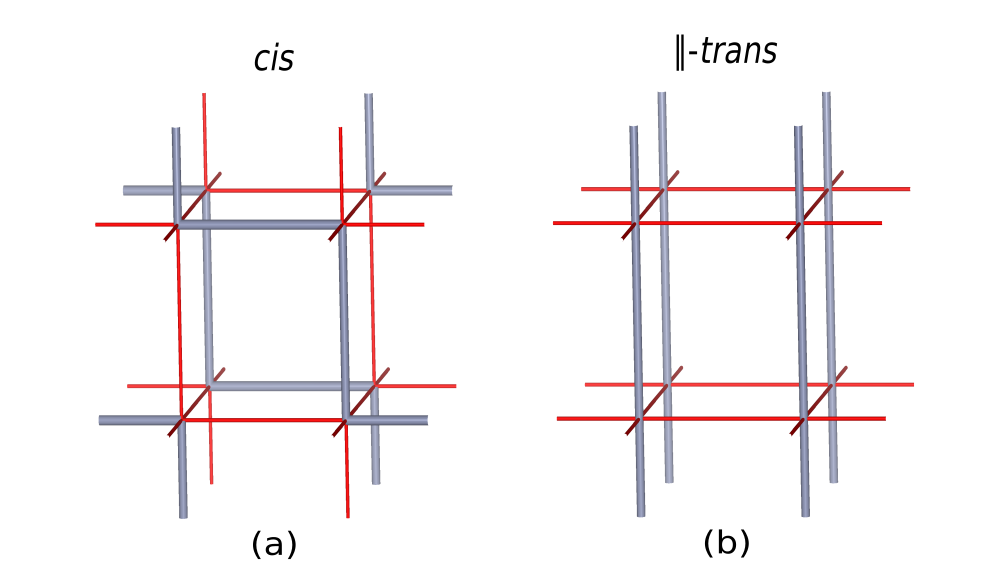}
\caption{Schematic representation of the two energetically most stable O/N orders: (a) \textit{cis} - N-Ti-N chains form a 2D network with 90\textdegree angles, (b) $\parallel$-\textit{trans} -
 180\textdegree N-Ti-N chains parallel to each other.
Color code: N-Ti-N=grey,O-Ti-O=red.}
\label{fig:anion-ordering}
\end{figure}

In the present work we compare eight different 2D anion orders. 3D anion arrangements were not considered as epitaxial strain will lead to structural anisotropy that favors arrangements of N-Ti-N bonds in a given plane or direction. These eight anion orders can be oriented along three directions each and epitaxial strain was applied in the three \{001\} planes, resulting in five structures per anion order after considering symmetries (see \SI{supporting information} for details). For all anion arrangements, we relaxed the out-of-plane lattice parameter and all internal coordinates for epitaxial strains between -5\% and +5\%. $\Gamma$-point phonons were computed to monitor the appearance of structural instabilities and displace the structures along unstable normal modes.

We found the two anion arrangements schematically shown in Fig. \ref{fig:anion-ordering} to be energetically favorable throughout the whole range of strain (see \SI{supporting information} for the remaining anion orders). In the \textit{cis} order (Fig. \ref{fig:anion-ordering}a), which is stable in the unstrained bulk, chains of N-Ti-N bonds contain 90\textdegree angles, whereas in the $\parallel$-\textit{trans} order (Fig. \ref{fig:anion-ordering}b) they form linear chains. To distinguish between these anion orders under strain, 
we introduce the following notation: $a_{s}^{b}$, where $a$ denotes the anion order (\textit{cis}/$\parallel$-\textit{trans}), $b$ the direction/plane along/in which the N-Ti-N bonds are aligned and $s$ the strain plane.
For example \textit{cis}$^{bc}_{ac}$ refers to an anion order as shown in Fig. \ref{fig:anion-ordering}a, where N-Ti-N chains run in b and c direction with epitaxial strain applied along a and c. 

\begin{figure*}
\centering
  \includegraphics[width=0.95\textwidth]{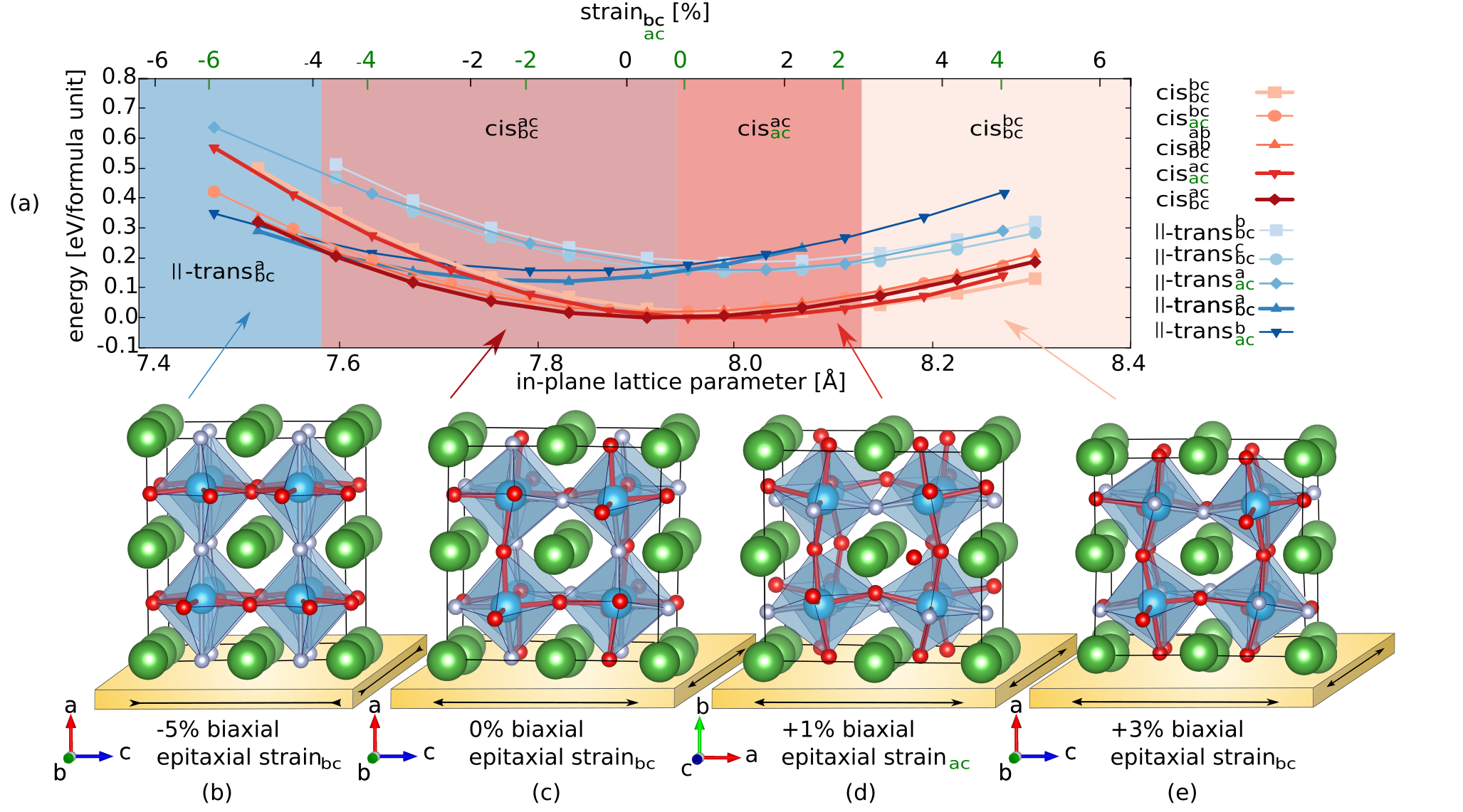}
  \caption{(a) Total energy as a function of the in-plane lattice parameter. The black and green strain axes at the top refer to the bc and ac area of the relaxed bulk structure respectively. 
Thermodynamically stable structures in the different strain regimes:
(b) $\parallel$-\textit{trans}$_{bc}^{a}$, stable for 
compressive strains\textsubscript{bc} smaller than 4\%; (c) \textit{cis}$^{ac}_{bc}$,  
stable between -4 and 0\% strain\textsubscript{bc}; (d) \textit{cis}$_{ac}^{ac}$, stable between 0 and 2\% strain\textsubscript{ac}; (e) \textit{cis}$_{bc}^{bc}$, stable for more than 3\% tensile strain\textsubscript{bc}.}
 \label{fig:all_energies}
\end{figure*}

In Fig. \ref{fig:all_energies}a we show the computed total energies as a function of the imposed in-plane lattice parameter for the two energetically relevant anion orders. The \textit{cis}-ordered structures are low in energy for the whole investigated range of strain, whereas the \textit{trans}-ordered structures are higher in energy at moderate strains. At the equilibrium lattice parameter, the energy difference between the \textit{cis} and the \textit{trans} order is around 0.2 eV per formula unit, which implies that the \textit{cis} order is favored even up to very high temperatures. Differently oriented \textit{cis} orders however have similar energies, in agreement with the experimentally observed absence of long-range order.

The experimentally observed bulk structure with N-Ti-N bonds along a and c (\textit{cis}$^{ac}$) is thermodynamically stable for strains from -4\% to +2\%, 
being applied either along ac (\textit{cis}$^{ac}_{ac}$, see Fig. \ref{fig:all_energies}d) or bc (\textit{cis}$^{ac}_{bc}$, see Fig. \ref{fig:all_energies}c).  
For tensile strains larger than 2\%, \textit{cis}$^{bc}_{bc}$, for which the N-Ti-N bonds lie along the elongated directions, becomes energetically 
more stable (see Fig. \ref{fig:all_energies}e).

Out of the five structures with a $\parallel$-\textit{trans} anion order, the two structures with a lower energy under compressive strain have N-Ti-N bonds along the out-of-plane direction ($\parallel$-\textit{trans}$_{ac}^{b}$ and $\parallel$-\textit{trans}$_{bc}^{a}$), 
whereas the three structures that show higher energies under compressive strain have the N-Ti-N bonds along one of the in-plane directions ($\parallel$-\textit{trans}$_{ac}^{a}$, $\parallel$-\textit{trans}$_{bc}^{c}$ and $\parallel$-\textit{trans}$_{bc}^{b}$). 
For compressive strain larger than 4\% $\parallel$-\textit{trans}$_{bc}^{a}$ (see Fig. \ref{fig:all_energies}b) with N-Ti-N bonds in 
the out-of-plane direction becomes energetically more stable than any of the \textit{cis}-type orders.

These observations can be explained by the larger ionic radius of N compared to O \cite{Shannon:1976vx}, 
which favors Ti-N bonds along elongated directions \footnote{in the perfect five-atom unit-cell we obtain Ti-N bonds that are 3\% longer than the Ti-O bonds}.
While this preferential alignment is possible for the \textit{cis} anion order under tensile strain, compressive strain favors \textit{trans} orders with N-Ti-N bonds along the out-of-plane direction.

Having established the anion order as a function of strain, we now turn to the polar properties of the thermodynamically stable structures. For charge carrier separation, only structures with a net polarization in the out-of-plane direction are technologically relevant as for in-plane polarization, carriers will have a tendency to recombine at domain walls. 
To reduce the computational cost of an initial screening of the polarization $\vec{p}$, we use a point-charge model with nominal ionic charges: 
\begin{equation}
\vec{p}=\sum_{i}\vec{x_i} q_i,
\end{equation}
where $\vec{x_i}$ is the position of atom $i$ and $q_i$ its nominal charge (La=+4, Ti=+3, O=-2, N=-3).
The polarization is a multivalued quantity - depending on the choice of unit cell and basis - that has to be corrected by an integer number
of polarization quanta $\vec{Q}$:
\begin{equation}
\vec{Q}=\frac{e}{V}\begin{pmatrix} a \\ b \\ c \\ \end{pmatrix},
\label{eq:quantum}
\end{equation}
where $e$ is the elementary charge, $V$ the volume of the unit cell and $a$, $b$ and $c$ are the lengths of the primitive basis vectors. The number of polarization quanta to be added/subtracted was determined relative to the non-polar transition state between the two polarization directions obtained using climbing-image nudged-elastic band calculations (see \SI{supporting information} for details). 

\begin{figure}
\centering
  \includegraphics[width=0.9\columnwidth]{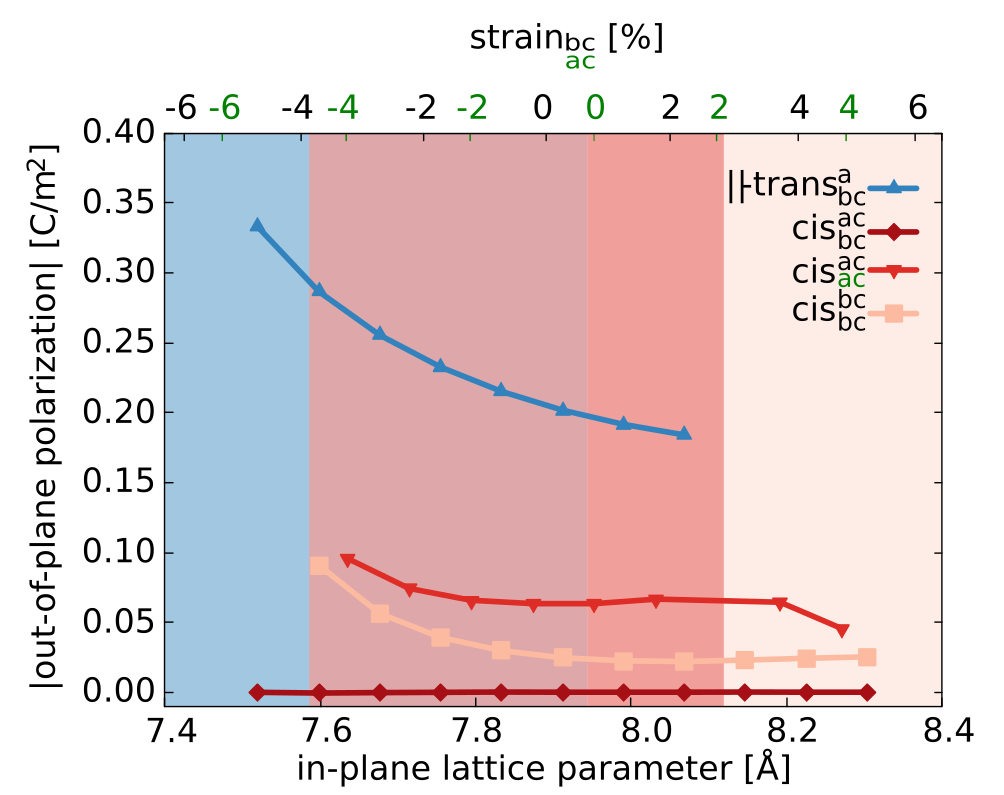}
  \caption{Out-of-plane polarization as a function of the in-plane lattice parameter for the four stable structures. The black and green strain axes at the top refer to the bc and ac area of the relaxed bulk structure respectively.} 
 \label{fig:polarization}
\end{figure}

As shown in Fig. \ref{fig:polarization}
we find, if allowed by symmetry, an increased polarization for directions elongated by strain, which agrees with the second-order Jahn-Teller picture.
 The structures $\parallel$-\textit{trans}$_{bc}^{a}$ and \textit{cis}$^{ac}_{ac}$ show a polarization in the out-of-plane direction without any polarization in the in-plane directions (in-plane components of the polarization are shown in the \SI{supporting information}). The \textit{cis}$^{bc}_{bc}$ structure also has a polarization in 
the out-of-plane direction, however combined with a high polarization along the two in-plane directions, which likely results in a smaller efficiency in 
suppressing electron-hole recombination. \textit{cis}$^{ac}_{bc}$ finally does not show any out-of-plane polarization, while one of the technologically less interesting in-plane directions increases in polarity for tensile strains.

Based on this initial screening we perform Berry-phase calculations of
the polarization for the $\parallel$-\textit{trans}$_{bc}^{a}$ structure at 5\% compressive strain and the \textit{cis}$^{ac}_{ac}$ structure at 1\% tensile strain, both having a non-zero polarization only in the out-of-plane direction.
These calculations yield polarizations of 0.59 C/m$^2$ and  0.13 C/m$^2$ respectively, showing that the electronic contribution captured in these calculations increases the polarization roughly a factor two compared to the point-charge model.

We estimate the coercive field required for switching the polarization using the following equation:
\begin{equation}
\epsilon_c = \frac{(4/3)^{(3/2)}E}{P \cdot V},
\label{eq:polarization}
\end{equation}
where $E$ is the switching barrier per formula unit, $P$ the polarization at the bottom of the double well potential  normalized per area perpendicular to the polarization direction and $V$
the volume per formula unit \cite{rabe2007physics, beckman2009ideal}.

\begin{figure}
\centering
  \includegraphics[width=\columnwidth]{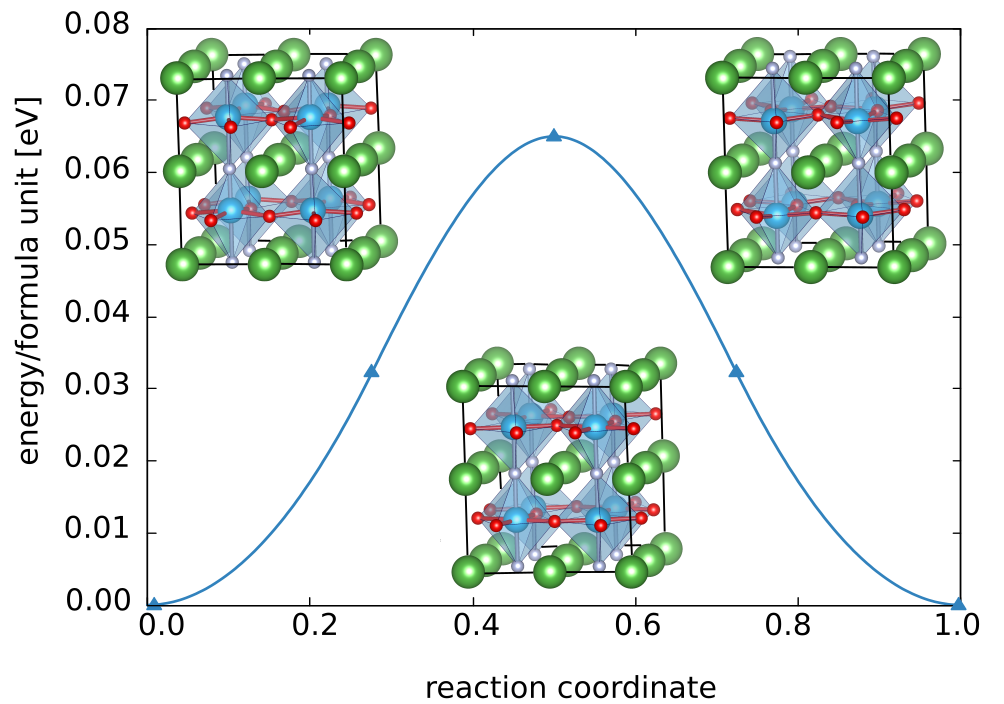}
  \caption{Polarization switching energy profile of $\parallel$-\textit{trans}$_{bc}^{a}$ at 5\% compressive strain.}
 \label{fig:neb}
\end{figure}

We evaluate the energy barrier between the two minimum-energy states of the double well via nudged elastic band calculations and find 0.065 eV and 0.23 eV for $\parallel$-\textit{trans}$_{bc}^{a}$ (see Fig. \ref{fig:neb}) and \textit{cis}$^{ac}_{ac}$ respectively. Using equation \ref{eq:polarization} and the Berry-phase polarizations we obtain coercive fields of 228 MV/m and 3476 MV/m per formula unit for $\parallel$-\textit{trans}$_{bc}^{a}$ and \textit{cis}$^{ac}_{ac}$ respectively. These results should however not be directly compared with experimentally measured coercive fields 
as the theoretical field correspond to the disappearance of the minority domain, whereas experimentally one measures the field at which domain walls become unpinned.
Fields calculated with this procedure are therefore likely to be significantly higher than experimental measurements. 
However, the calculated value for $\parallel$-\textit{trans}$_{bc}^{a}$ at -5\% strain is of the same order as other computationally obtained values for well-known switchable ferroelectrics such as PbTiO\textsubscript{3} \cite{beckman2009ideal}. Based on this comparison, we assume that LaTiO$_2$N films at compressive strains larger than 4\% should also be switchable with fields comparable to those of conventional ferroelectrics.

In summary, we have shown that for compressive epitaxial strain larger than 4\% LaTiO\textsubscript{2}N adopts a \textit{trans}-type anion order, which has spontaneous out-of-plane polarization. 
The computed theoretical coercive field of 228 MV/m is in the same range as computations for other well-known switchable ferroelectrics such as PbTiO\textsubscript{3}. 
We therefore expect that switching the polarization in compressively strained LaTiO\textsubscript{2}N films should be possible with reasonable electric fields. Our findings highlight the possibility to engineer ferroelectricity in materials that already possess an electronic structure suitable for visible-light photocatalysis. Ferroelectric photocatalyst will enhance charge-carrier separation and could, when switched, enable alternative reaction pathways for water oxidation that no longer obey Sabatier's principle.

\section{Acknowledgements}

This research was funded by the SNF Professorship Grant PP00P2\_157615. Calculations were performed on UBELIX (http://www.id.unibe.ch/hpc), the HPC cluster at the University of Bern.

\bibliography{bib}

\end{document}